# Multi-wavelength magnetic coding of helical luminescence in ferromagnetic 2D layered CrI$_3$

Bo Peng[1,2,*], Zhiyong Chen[1,2], Yue Li[1,2], Zhen Liu[1,2], Difei Liang[1,2,*], Longjiang Deng[1,2]

[1]National Engineering Research Center of Electromagnetic Radiation Control Materials, School of Electronic Science and Engineering, University of Electronic Science and Technology of China, Chengdu, 611731, China

[2]State Key Laboratory of Electronic Thin Films and Integrated Devices, University of Electronic Science and Technology of China, Chengdu, 611731, China

*Correspondence author's e-mail: bo_peng@uestc.edu.cn; dfliang@uestc.edu.cn

**Abstract:** Two-dimensional (2D) van der Waals (vdW) ferromagnets have opened new avenues for manipulating spin at the limits of single or few atomic layers, and for creating unique magneto-exciton devices through the coupling of long-range ferromagnetic (FM) orders and excitons. However, 2D vdW ferromagnets explored so far have rarely possessed exciton behaviors; to date, FM $CrI_3$ have been recently revealed to show ligand-field photoluminescence correlated with FM ordering, but typically with a broad emission peak. Alternatively, many-body excitons have been observed in antiferromagnetic (AFM) $NiPS_3$, but the coupling of excitons with AFM orders is exponentially more difficult, owing to extremely high coercivity. Here, we report a straightforward approach to realize strong coupling of narrow helical emission and FM orders at a low magnetic field in $CrI_3$ through a relatively simple microsphere cavity. We show that the resonant whispering-gallery-modes (WGM) of $SiO_2$ microspheres give rising to a series of strong oscillation helical emissions with a full width at half-maximum (FWHM) of ~5 nm under continuous wave excitation. Reversible magnetic control and coding of helical luminescence with multiwavelength is realized in the range of 950-1100 nm. This work enables plenty of opportunities for creating magnetic encoding lasing for photonic integrated chips.

The emergence of long-range ferromagnetic orders in 2D vdW FM material has provided a new avenue for creating on-chip lasers, isolators and modulators for silicon photonics toward information processing and transmission through magneto-exciton coupling (Deng et al., 2018; Gong et al., 2017; Huang et al., 2017; Liu et al., 2021; Liu et al., 2020; Mak et al., 2019; Song et al., 2018; Sun et al., 2019). The magnetic orders can be switched between FM and AFM states by magnetic field (Klein et al., 2018; Zhong et al., 2017), electric gating (Huang et al., 2018; Jiang et al., 2018a; Jiang et al., 2018d; Zhang et al., 2020), electrostatic doping (Jiang et al., 2018b; Zheng et al., 2020) and hydrostatic pressure (Li et al., 2019; Song et al., 2019), which empower a pivotal foundation for realizing magnetic and electric control of magneto-optical coupling for information transfer. 2D vdW bilayer $CrI_3$ FM insulators used as spin valve in a magnetic tunneling junction device have highlighted flexibilities for encoding information and data storage (Jiang et al., 2018c; Klein et al., 2018; Song et al., 2018). To date, diverse 2D vdW FM and AFM metal, insulators and even semiconductors have been explored (Liu et al., 2021). FM $Cr_2Ge_2Te_6$ few-layer have demonstrated a p-type semiconducting behavior and gate-controllable magnetism (Jiang et al., 2018b; Verzhbitskiy et al., 2020); few-layer $Fe_3GeTe_2$ FM metal have shown high Curie temperature ($T_c$) above room temperature through ionic/protonic gating, patterning and interfacial exchange coupling (Deng et al., 2018; Li et al., 2018; May et al., 2019; Wang et al., 2020a); magnetic topological insulator $MnBi_2Te_4$ have recently been reported to present layer-dependent magnetic phase and spin-flop behaviors (Deng et al., 2020; Yang et al., 2021). Alternatively, air-stable metallic 2D vdW $CrTe_2$ and $CrSe_2$ with thickness-dependent magnetic orders have been directly synthesized by chemical vapor deposition (CVD) methods (Li et al., 2021; Meng et al., 2021). However, thus far, few 2D vdW magnets have been revealed luminescence and exciton behaviors, and none examining magnetic coding of photoluminescence (PL), despite the discovery of luminescence in FM $CrI_3$ and AFM $NiPS_3$ (Hwangbo et al., 2021; Kang et al., 2020; Seyler et al., 2017; Wang et al., 2021). Moreover, the PL of $CrI_3$ derives from a ligand-field-allowed d-d transition, which naturally lead to a very broad PL peak with a FWHM of 100-200 nm owing to strong vibrionic coupling (Seyler et al., 2017), extremely limiting the potential for photonic devices; in addition, it is difficult to achieve magnetic/electric control of spin in AFM $NiPS_3$, due to a high critical field up to 10 T at least (Wang et al., 2021).

With the rapid development of digitalization and semiconductor optoelectronics technology, various fields have an increasing demand for on-chip lasers, so the research on high-performance and miniaturized on-chip lasers is of great significance (Gao et al., 2020; Kyeremateng et al., 2017). On-chip lasers have broad application prospects, such as a light source for on-chip integrated photonic chips, laser scanning and ranging or flexible displays. Traditional directly modulated lasers (DML) have been achieved by injecting the pump current to the gain medium, and the modulation speed is affected by its volume of the active region and relaxation oscillation frequency (Dong et al., 2018; Kobayashi et al., 2013). Alternatively, topological lasers with single mode lasing and robustness against defects have been realized by coupling of photonic gain with topological structure (Bandres et al., 2018). Additionally, atomic-thin lasers using transition-metal dichalcogenides (TMDs) with strong exciton emission as optical gain medium have been recently reported (Li et al., 2017; Salehzadeh et al., 2015a; Wu et al., 2015; Ye et al., 2015b). However, thus far, small-volume atomic-thin lasers with intrinsic magnetism in the near-infrared (NIR) region have been still elusive, which provide numerous opportunities to realized magnetic- field controlled DML. Moreover, the NIR laser is more challenging duo to the lack of gain medium with strong NIR emission.

Here, we report the realization of magnetic control and coding of NIR WGM emission through coupling of the luminescence of 2D ferromagnetic $CrI_3$ and $SiO_2$ microsphere cavities. Circularly polarized WGM PL features in $CrI_3$/microsphere are coupled with $CrI_3$ magnetic orders. The right-handed (left-handed) polarized WGM PL are dominated in a positive (negative) magnetic field of +1 T (-1 T). Multi-wavelength encoding is realized through repeatedly manipulating helicity of each WGM PL peak by magnetic field. This work opens the door to creating new atomically thin 2D magneto-optical devices for photonic integrated circuit and on-chip optical system.

## Results

**Coupling of magnetic orders and PL in $CrI_3$.** In 2D FM $CrI_3$ monolayer, $Cr^{3+}$ ions in each layer are coordinated by six nonmagnetic $I^-$ ions to form an octahedron (Guo et al., 2020), which shares edges to build a honeycomb network (Fig. 1A). Monolayer $CrI_3$ is a Ising ferromagnetism arising from the Cr-I-Cr superexchange interaction, which is described by the Ising spin Hamiltonian, $H = -(1/2)\sum(J_{xy} S_{i(x)} S_{j(x)} + J_{xy} S_{i(y)} S_{j(y)} + J_z S_{i(z)} S_{j(z)})$, where $S_{i(x, y, z)}$ and $S_{j(x, y, z)}$ are the spin along the $x, y, z$ direction of $Cr^{3+}$ at $i(j)$ sites;

$J_{xy}$ and $J_z$ are the exchange coupling term of in-plane and out-of-plane spin components and $J_z > J_{xy} > 0$ in Ising FM model. In few-layer CrI$_3$, intra-atomic *d-d* transitions and charge-transfer transitions owing to ligand field lead to a broad layer-dependent PL emission at ~1020 nm with a FWHM of ~100 nm (Supplementary information, Fig. S1) (Seyler et al., 2017). PL intensities show strong layer dependence and significantly increase as increasing layers (Seyler et al., 2017), therefore, thick CrI$_3$ flakes and bulk were choose in our experiments. The helicity of the absorption and associated PL emission is correlated with the spin of the electrons at the ground states. Spin states are tied to magnetic orders of CrI$_3$ few-layers. Thus, in an out-of-plane positive magnetic field, the upper spin states are dominated and absorb right-hand polarized light, consequentially mostly giving right-hand polarized PL emission (Fig. 1B); when the magnetic field turns downward, the spins of electrons are reversed, and mainly lead to left-hand polarized PL (Supplementary information, Fig. S1). But the poor and broad PL features restrict the exploring on magneto-exciton coupling and creating photonic devices with unique functions. Optical microcavities have been widely applied to enhance light-matter interaction and suppress FWHM for realizing strong emission and lasing (Salehzadeh et al., 2015b; Schwarz et al., 2014; Wu et al., 2014; Ye et al., 2015a). SiO$_2$ microsphere as high-quality WGM microcavities with high Q-factor of ~$10^8$ can significantly enhance the coupling between gain medium and optical cavity modes (Mi et al., 2017; Zhao et al., 2018). Thus, SiO$_2$ microsphere cavities can be coupled with FM 2D layered CrI$_3$ to create narrow and strong WGM oscillation PL peaks. The coupling between circularly polarized PL emission and SiO$_2$ spherical microcavities obey the law of conservation of angular momentum, thus, the PL helicity associated with magnetic orders are preserved. In a positive magnetic field, right-hand polarization-resolved WGM oscillation PL originating from spin-up states are dominated, and vice versa (Fig. 1C). Polarization-resolved Raman spectra show that thick layered CrI$_3$ bulk sample adopts a rhombohedral structure at 10 K. Raman shift of the two-fold degenerated $E_g$ mode at ~107 cm$^{-1}$ are independent of the polarization angle (Fig. 2A and Fig. S2), validating the high-symmetry rhombohedral phase (Li et al., 2019; Song et al., 2019). Reflective magnetic circular dichroism (RMCD) was used to probe the long-range FM orders. A 633 nm continuous-wave laser was applied as an excitation source with the light spot size of ~2 μm and 2 μW. Figure 2B and C show the typical RMCD curve of bare CrI$_3$ and CrI$_3$ beneath SiO$_2$ sphere. Striking hysteresis loops are observed on both bare CrI$_3$ and CrI$_3$ beneath SiO$_2$ sphere below 55 K, indicating a $T_c$ of ~50 K and the SiO$_2$ microsphere cavities do not influence the $T_c$ of CrI$_3$.

**Magnetic control of WGM PL emission of CrI$_3$.** To obtain strong enhancements, SiO$_2$ microspheres with 10 and 12.5 μm diameter were transferred onto mechanically exfoliated thick CrI$_3$ flakes (~32 nm) and bulk (~55 nm) through polydimethylsiloxane (PDMS) films (Supplementary information, Fig. S2). The Q factors of SiO$_2$ spherical cavities were firstly estimated on MoS$_2$/SiO$_2$ samples. The Q factors of microsphere cavities are estimated from Q=$\lambda/\delta\lambda$, where $\delta\lambda$ is the FWHM of the WGMs, and the Q factor for the TM$_{1,61}$ mode (~633 nm) can reach about 1055(supplementary information, Fig. S3). As shown in Figure S3B, the PL intensity increases sharply with increasing excitation light power, and the lasing behavior at the $TM_{1,61}$ mode (~633 nm) is observed. Figure S3C shows that PL intensity of $MoS_2/SiO_2$ and $MoS_2$ as a function of laser power. The PL intensities of $MoS_2$ shows a linear dependence on the excitation light power, and $TM_{1,61}$ mode shows a kink indicating the onset of superlinear emission. In Figure S3D, the FWHM of $TM_{1,61}$ mode decreases from 0.9 to 0.6 nm, the linewidth narrowing effect further confirms the appearance of lasing. Room-temperature lasing from $MoS_2$/sphere validate that Q-factor of $SiO_2$ is high enough for WGM lasing (Zhao et al., 2018). WGM PL peaks in the range of ~900-1100 nm are unambiguously observed from $CrI_3$ beneath 10 and 12.5 μm $SiO_2$ spheres under continuous wave 633 nm excitation with 1 mW (Supplementary information, Fig. S4). $SiO_2$ microsphere optical cavities enhance PL intensity of $CrI_3$ by ~2.7 times as compared to that from the same but bare $CrI_3$. The excitation area can be effectively reduced to enhance the coupling between the gain region and the optical mode due to the lens effect of the microspheres (Mi et al., 2017; Zhao et al., 2018), which increases the spontaneous emission efficiency and PL intensity of $CrI_3$. In stark contrast, no any oscillation peak is detected in bare $CrI_3$ sample. Theoretical calculations demonstrate that the strong WGM PL are attributed to the first-order TM modes ($TM_1$), as indicated by blue line (Supplementary information, Fig. S4).

To study the WGM PL in the microsphere cavity, power-dependent PL spectra were recorded on $CrI_3$/microsphere vertical light-emitting gain structure in the range from 10 μW to 1.5 mW with a 633 nm laser excitation source at 10 K. The lensing and screening effects of the microsphere cavity increase the excitation efficiency of $CrI_3$, the equally-spaced oscillation peaks can be distinguished, even the excitation power is as low as about 10 μW (Supplementary information, Fig. S5 and S6). The WGM PL intensities linearly increase with laser power increasing. When the laser power is increased to 1.5 mW, the oscillation peaks intensity increases sharply and the peak width is narrowed. The spacing between two adjacent resonant modes remains constant with increasing laser power, indicating that the WGM PL originating from the same resonant modes of the sphere cavities. The PL intensities of bare $CrI_3$ samples also linearly increase with the increase of laser power, but smaller than WGM PL intensity. The integrated PL intensities of $TM_{1,36}$, $TM_{1,37}$, $TM_{1,47}$ and $TM_{1,48}$ modes of $CrI_3$/microsphere with 10 and 12.5 μm $SiO_2$ spheres remains linear to the excitation power density. With further increasing laser power, the WGM oscillation peak intensity of the sample achieve maximum at 2.6 mW and drastically decrease at 3.0 mW due to laser burning of $CrI_3$ flakes. But no transition from spontaneous emission to amplified spontaneous emission was observed (Supplementary information, Fig. S5 and S6). This may be because the gain of $CrI_3$ flakes is not strong enough and they are also easy to burn out under high power, resulting in no super linear response. The PL intensity of thin $CrI_3$ was strongly dependent on the number of layers, and decrease with decreasing thickness of $CrI_3$. Even no WGM peaks are observed and only PL enhancement behaviors take place if the gain of $CrI_3$ is decreased by applying thin $CrI_3$ flakes (Supplementary information, Fig. S7). WGM PL features are nearly same in 10 and 12.5 μm $SiO_2$ microsphere cavities. Therefore, we focus on 10 μm $SiO_2$ microcavities to further study the magnetic control and coding of WGM PL.

WGM PL of CrI$_3$/microsphere can be manipulated by an out-of-plane magnetic field. Figure 3 presents the circularly polarization-resolved WGM PL features from CrI$_3$/microsphere at 10 K, at magnetic field +1, 0, -1 and 0 T, respectively. We define the right-hand (left-hand) circularly polarized PL spectra upon excitation by right-hand (left-hand) circularly polarized laser as RR (LL). The intensity of circularly polarized WGM PL is correlated to magnetic orders of CrI$_3$. In a magnetic field of +1 T, the FM orders are spin-up, leading to that the spin-up states are dominated, therefore, RR WGM PL is stronger than LL WGM PL (Fig. 3A and E). As the magnetic field is lowered to 0 T, the CrI$_3$ show an anti-ferromagnetic behavior, and spin-up and spin-down states are nearly equal (Fig. 3B and E). When magnetic field is further reversed to -1 T, spin states are also inversed, giving rise to opposite helicity and exhibiting a stronger LL component than RR (Fig. 3C and E). When magnetic field return back to 0 and +1 T, the WGM PL is also recovered (Fig. 3D and A). The corresponding RMCD results clearly show the associated FM and AFM orders, which consist with the magnetic control behaviors of circularly polarization-resolved WGM PL. It should be noted that the intensity variations of right-handed and left-handed polarized PL of the CrI$_3$/sphere are lower than the same pure CrI$_3$ flakes (Fig. 3 and supplementary information, Fig. S1). This probably results from the decrease of magnetic anisotropy and long-range ferromagnetic orders induced by tensile strains, which are formed during the transferring SiO$_2$ sphere onto CrI$_3$ by a PPMA fixed-point transfer method. Previous reports have demonstrated that tensile strains lead to an increase of bandgap associated with unchanged valence band maximum (VBM) and higher conduction band minimum

(CBM) (Webster and Yan, 2018; Zhang et al., 2015). Fig. S7B shows that the PL peak is blue-shifted to a high energy, which indicates that tensile strains are formed on $CrI_3$ flakes. Tensile strains lead to the increase of the Cr-I-Cr angle and bond length, which result in decreasing magnetic anisotropy energies and ferromagnetic orders of $CrI_3$. The helicity of PL emission of $CrI_3$ is strongly tied to magnetic orders. Therefore, the intensity variations of circularly polarized PL of $CrI_3$/sphere decrease.

**Multiwavelength magnetic encoding**. The WGM oscillation modes of sphere cavities result in a series of narrow PL at multiwavelength. We studied the multiwavelength magnetic encoding of circularly polarized PL through recycling a loop of magnetic field between +1 and -1 T. Magnetic field loop is set from +1 to 0 T and inversely increase to -1 T, and return back to +1 T (Fig. 4A). The PL intensity differences between LL and RR WGM PL, defined as $\Delta I$, show striking narrow oscillation peaks at 976, 995, 1020, 1045 and 1070 nm with FWHM between 5 and 10 nm (Fig. 4B), which show an obvious magnetic-field dependent behavior. $\Delta I$ at the five distinct wavelengths are opposite at +1 and -1 T, and much larger than that at 0 T, which provide a rich platform for tri-state encoding through on/off switching of magnetic field. Figure 4b show ten recycling of $\Delta I$ for multiwavelength magnetic encoding. The $\Delta I$ features at 1020±5, 1070±5, 1045±5 and 995±5 nm were extracted. Each wavelength presents +1, 0 and -1 tri-states at magnetic field of +1, 0 and -1T. The magnetic control behaviors of tri-states are entirely correlated to the magnetic field and display the same shape with magnetic field setups, as shown in Fig. 4C and Fig. 5. The helicity of $CrI_3$/microsphere WGM oscillation peaks are determined by magnetic orders of $CrI_3$ and controlled by magnetic field, which also can be used to achieve magnetic encoding function. Similar magnetic-field control behaviors have been achieved on different samples (Fig. 4B and Fig. 3), although the PL intensity variations is different, which strongly validate that the $SiO_2$ spherical cavities can generally lead to WGM peaks for $CrI_3$ ferromagnetism and reversible magnetic control and coding of helical luminescence in the $CrI_3$/microsphere is reproducible. The degree of polarization (DOP) from $CrI_3$/microsphere and $CrI_3$ is given by DOP = $(I_{RR} - I_{LL})/ (I_{LL} + I_{RR})$, where $I_{RR}$ and $I_{LL}$ are the intensity of right-hand and left-hand circularly polarized PL (Li et al., 2020; Peng et al., 2017; Wang et al., 2020b). When a magnetic field of +1 T is applied, DOP of the WGM PL is around +10%, while the DOP is inversely changed to -10% in a -1 T magnetic field, due to the reversal of the magnetization of $CrI_3$ (Fig. 6). Magnetic fields manipulate the DOP of each oscillation peak in ten recycling, clearly validating multiwavelength magnetic encoding behaviors.

**Discussion**

In summary, we demonstrated magnetic control of NIR helical WGM PL and realized tri-state encoding at different wavelength. The lensing and screening effects of $SiO_2$ microspheres increase the spontaneous emission efficiency of $CrI_3$ to create unambiguous WGM oscillation peaks which is coupled with magnetic orders of $CrI_3$. Magnetic fields manipulate and control each WGM peak, and the PL helicity and intensity difference of every WGM peak are applied to realize multiwavelength encoding through recycling magnetic fields. This work provides promising opportunities for creating magneto-optical devices and integrated photonic lasing with magnetic encoding function through FM 2D materials microcavity.

## Methods

**Sample preparation.**

CrI$_3$ flakes were mechanically exfoliated from bulk crystal onto polydimethylsiloxane films and then transferred onto SiO$_2$/Si substrates. The commercial SiO$_2$ microspheres solution were dropped onto another SiO$_2$/Si substrates, which were heated on a heating table to make the solvent completely volatilize. Next, SiO$_2$ microspheres were directly transferred onto CrI$_3$ flakes. The samples were loaded into cold head for optical measurements in glove box.

**Optical measurements.**

The Raman and PL signal were recorded by Witec Alpha 300R Plus confocal Raman microscope, coupled with a closed-cycle He optical cryostat (10 K) and a 7 T magnetic field. The excitation laser of 633 nm was 2 mW and the integration time was 15 s. A 633 nm HeNe laser with ~2 μW was coupled to the Witec Raman system through free optical path for RMCD measurements, which was modulated by photoelastic modulator (PEM, $f_{PEM}$ = 50 KHz) and focused onto samples by a long working distance 50× objective (NA = 0.45). The reflected beams were collected by the same objective, passed through a non-polarizing beamsplitter cube into an APD photodetector.

**Data availability**. All relevant data are available from the corresponding authors on request.

## Reference

Bandres, M.A., Wittek, S., Harari, G., Parto, M., Ren, J., Segev, M., Christodoulides,

D.N., and Khajavikhan, M. (2018). Topological insulator laser: Experiments. Science *359* http://dx.doi.org/10.1126/science.aar4005.

Deng, Y., Yu, Y., Shi, M.Z., Guo, Z., Xu, Z., Wang, J., Chen, X.H., and Zhang, Y. (2020). Quantum anomalous Hall effect in intrinsic magnetic topological insulator $MnBi_2Te_4$. Science *367*, 895-900. http://dx.doi.org/10.1126/science.aax8156.

Deng, Y., Yu, Y., Song, Y., Zhang, J., Wang, N.Z., Sun, Z., Yi, Y., Wu, Y.Z., Wu, S., Zhu, J., et al. (2018). Gate-tunable room-temperature ferromagnetism in two-dimensional $Fe_3GeTe_2$. Nature *563*, 94-99. http://dx.doi.org/10.1038/s41586-018-0626-9.

Dong, P., Maho, A., Brenot, R., Chen, Y.-K., and Melikyan, A. (2018). Directly Reflectivity Modulated Laser. Journal of Lightwave Technology *36*, 1255-1261. http://dx.doi.org/10.1109/jlt.2018.2791363.

Gao, Y., Lo, J.-C., Lee, S., Patel, R., Zhu, L., Nee, J., Tsou, D., Carney, R., and Sun, J. (2020). High-Power, Narrow-Linewidth, Miniaturized Silicon Photonic Tunable Laser With Accurate Frequency Control. Journal of Lightwave Technology *38*, 265-271. http://dx.doi.org/10.1109/jlt.2019.2940589.

Gong, C., Li, L., Li, Z., Ji, H., Stern, A., Xia, Y., Cao, T., Bao, W., Wang, C., Wang, Y., et al. (2017). Discovery of intrinsic ferromagnetism in two-dimensional van der Waals crystals. Nature *546*, 265–269. http://dx.doi.org/10.1038/nature22060.

Guo, K., Deng, B., Liu, Z., Gao, C., Shi, Z., Bi, L., Zhang, L., Lu, H., Zhou, P., Zhang, L., et al. (2020). Layer dependence of stacking order in nonencapsulated few-layer $CrI_3$. Sci China Mater *63*, 413-420. http://dx.doi.org/10.1007/s40843-019-1214-y.

Huang, B., Clark, G., Klein, D.R., MacNeill, D., Navarro-Moratalla, E., Seyler, K.L.,

Wilson, N., McGuire, M.A., Cobden, D.H., Xiao, D., et al. (2018). Electrical control of 2D magnetism in bilayer $CrI_3$. Nat Nanotechnol *13*, 544-548. http://dx.doi.org/10.1038/s41565-018-0121-3.

Huang, B., Clark, G., Navarro-Moratalla, E., Klein, D.R., Cheng, R., Seyler, K.L., Zhong, D., Schmidgall, E., McGuire, M.A., Cobden, D.H., et al. (2017). Layer-dependent ferromagnetism in a van der Waals crystal down to the monolayer limit. Nature *546*, 270-273. http://dx.doi.org/10.1038/nature22391.

Hwangbo, K., Zhang, Q., Jiang, Q., Wang, Y., Fonseca, J., Wang, C., Diederich, G.M., Gamelin, D.R., Xiao, D., Chu, J.-H., et al. (2021). Highly anisotropic excitons and multiple phonon bound states in a van der Waals antiferromagnetic insulator. Nat Nanotechnol *16*, 655-660. http://dx.doi.org/10.1038/s41565-021-00873-9.

Jiang, S., Li, L., Wang, Z., Mak, K.F., and Shan, J. (2018a). Controlling magnetism in 2D $CrI_3$ by electrostatic doping. Nat Nanotechnol *13*, 549-553. http://dx.doi.org/10.1038/s41565-018-0135-x.

Jiang, S., Li, L., Wang, Z., Mak, K.F., and Shan, J. (2018b). Controlling magnetism in 2D $CrI_3$ by electrostatic doping. Nat Nanotechnol *13*, 549-553. http://dx.doi.org/10.1038/s41565-018-0135-x.

Jiang, S., Li, L., Wang, Z., Mak, K.F., and Shan, J. (2018c). Controlling magnetism in 2D $CrI_3$ by electrostatic doping. Nat Nanotechnol *13*, 549-553. http://dx.doi.org/10.1038/s41565-018-0135-x.

Jiang, S., Shan, J., and Mak, K.F. (2018d). Electric-field switching of two-dimensional van der Waals magnets. Nat Mater *17*, 406-410 http://dx.doi.org/10.1038/s41563-018-

0040-6.

Kang, S., Kim, K., Kim, B.H., Kim, J., Sim, K.I., Lee, J.-U., Lee, S., Park, K., Yun, S., Kim, T., et al. (2020). Coherent many-body exciton in van der Waals antiferromagnet NiPS$_3$. Nature *583*, 785-789. http://dx.doi.org/10.1038/s41586-020-2520-5.

Klein, D.R., MacNeill, D., Lado, J.L., Soriano, D., Navarro-Moratalla, E., Watanabe, K., Taniguchi, T., Manni, S., Canfield, P., Fernández-Rossier, J., et al. (2018). Probing magnetism in 2D van der Waals crystalline insulators via electron tunneling. Science *360*, 1218-1222. http://dx.doi.org/10.1126/science.aar3617.

Kobayashi, W., Ito, T., Yamanaka, T., Fujisawa, T., Shibata, Y., Kurosaki, T., Kohtoku, M., Tadokoro, T., and Sanjoh, H. (2013). 50-Gb/s Direct Modulation of a 1.3-μm InGaAlAs-Based DFB Laser With a Ridge Waveguide Structure. IEEE Journal of Selected Topics in Quantum Electronics *19*, 1500908-1500908. http://dx.doi.org/10.1109/jstqe.2013.2238509.

Kyeremateng, N.A., Brousse, T., and Pech, D. (2017). Microsupercapacitors as miniaturized energy-storage components for on-chip electronics. Nat Nanotechnol *12*, 7-15. http://dx.doi.org/10.1038/nnano.2016.196.

Li, B., Wan, Z., Wang, C., Chen, P., Huang, B., Cheng, X., Qian, Q., Li, J., Zhang, Z., Sun, G., et al. (2021). Van der Waals epitaxial growth of air-stable CrSe$_2$ nanosheets with thickness-tunable magnetic order. Nat Mater http://dx.doi.org/10.1038/s41563-021-00927-2.

Li, Q., Yang, M., Gong, C., Chopdekar, R.V., N'Diaye, A.T., Turner, J., Chen, G., Scholl, A., Shafer, P., Arenholz, E., et al. (2018). Patterning-Induced Ferromagnetism of Fe$_3$GeTe$_2$ van der Waals Materials beyond Room Temperature. Nano Letters *18*, 5974-5980. http://dx.doi.org/10.1021/acs.nanolett.8b02806.

Li, Q., Zhao, X., Deng, L., Shi, Z., Liu, S., Wei, Q., Zhang, L., Cheng, Y., Zhang, L., Lu, H., et al. (2020). Enhanced valley Zeeman splitting in Fe-doped monolayer MoS$_2$. ACS Nano *14*, 4636-4645. http://dx.doi.org/10.1021/acsnano.0c00291.

Li, T., Jiang, S., Sivadas, N., Wang, Z., Xu, Y., Weber, D., Goldberger, J.E., Watanabe, K., Taniguchi, T., Fennie, C.J., et al. (2019). Pressure-controlled interlayer magnetism in atomically thin CrI$_3$. Nat Mater *18*, 1303-1308. http://dx.doi.org/10.1038/s41563-019-0506-1.

Li, Y., Zhang, J., Huang, D., Sun, H., Fan, F., Feng, J., Wang, Z., and Ning, C.Z. (2017). Room-temperature continuous-wave lasing from monolayer molybdenum ditelluride integrated with a silicon nanobeam cavity. Nat Nanotechnol *12*, 987-992. http://dx.doi.org/10.1038/nnano.2017.128.

Liu, Z., Deng, L., and Peng, B. (2021). Ferromagnetic and ferroelectric two-dimensional materials for memory application. Nano Res *14*, 1802-1813. http://dx.doi.org/10.1007/s12274-020-2860-3.

Liu, Z., Guo, K., Hu, G., Shi, Z., Li, Y., Zhang, L., Chen, H., Zhang, L., Zhou, P., Lu, H., et al. (2020). Observation of nonreciprocal magnetophonon effect in nonencapsulated few-layered CrI$_3$. Sci Adv *6*, eabc7628. http://dx.doi.org/10.1126/sciadv.abc7628.

Mak, K.F., Shan, J., and Ralph, D.C. (2019). Probing and controlling magnetic states in 2D layered magnetic materials. Nat Rev Phys *1*, 646-661.

http://dx.doi.org/10.1038/s42254-019-0110-y.

May, A.F., Ovchinnikov, D., Zheng, Q., Hermann, R., Calder, S., Huang, B., Fei, Z., Liu, Y., Xu, X., and McGuire, M.A. (2019). Ferromagnetism Near Room Temperature in the Cleavable van der Waals Crystal $Fe_5GeTe_2$. ACS Nano *13*, 4436-4442. http://dx.doi.org/10.1021/acsnano.8b09660.

Meng, L., Zhou, Z., Xu, M., Yang, S., Si, K., Liu, L., Wang, X., Jiang, H., Li, B., Qin, P., et al. (2021). Anomalous thickness dependence of Curie temperature in air-stable two-dimensional ferromagnetic $1T-CrTe_2$ grown by chemical vapor deposition. Nat Commun *12*, 809. http://dx.doi.org/10.1038/s41467-021-21072-z.

Mi, Y., Zhang, Z., Zhao, L., Zhang, S., Chen, J., Ji, Q., Shi, J., Zhou, X., Wang, R., Shi, J., et al. (2017). Tuning excitonic properties of monolayer $MoS_2$ with microsphere cavity by high-throughput chemical vapor deposition method. Small *13*, 1701694. http://dx.doi.org/10.1002/smll.201701694.

Peng, B., Li, Q., Liang, X., Song, P., Li, J., He, K., Fu, D., Li, Y., Shen, C., Wang, H., et al. (2017). Valley polarization of trions and magnetoresistance in heterostructures of $MoS_2$ and yttrium iron garnet. ACS Nano *11*, 12257-12265. http://dx.doi.org/10.1021/acsnano.7b05743.

Salehzadeh, O., Djavid, M., Tran, N.H., Shih, I., and Mi, Z. (2015a). Optically Pumped Two-Dimensional MoS2 Lasers Operating at Room-Temperature. Nano Lett *15*, 5302-5306. http://dx.doi.org/10.1021/acs.nanolett.5b01665.

Salehzadeh, O., Djavid, M., Tran, N.H., Shih, I., and Mi, Z. (2015b). Optically Pumped Two-Dimensional $MoS_2$ Lasers Operating at Room-Temperature. Nano Lett *15*, 5302-

5306. http://dx.doi.org/10.1021/acs.nanolett.5b01665.

Schwarz, S., Dufferwiel, S., Walker, P.M., Withers, F., Trichet, A.A.P., Sich, M., Li, F., Chekhovich, E.A., Borisenko, D.N., Kolesnikov, N.N., et al. (2014). Two-dimensional metal-chalcogenide films in tunable optical microcavities. Nano Lett *14*, 7003-7008. http://dx.doi.org/10.1021/nl503312x.

Seyler, K.L., Zhong, D., Klein, D.R., Gao, S., Zhang, X., Huang, B., Navarro-Moratalla, E., Yang, L., Cobden, D.H., McGuire, M.A., et al. (2017). Ligand-field helical luminescence in a 2D ferromagnetic insulator. Nature Physics *14*, 277-281. http://dx.doi.org/10.1038/s41567-017-0006-7.

Song, T., Cai, X., Tu, M.W.-Y., Zhang, X., Huang, B., Wilson, N.P., Seyler, K.L., Zhu, L., Taniguchi, T., Watanabe, K., et al. (2018). Giant tunneling magnetoresistance in spin-filter van der Waals heterostructures. Science *360*, 1214-1218. http://dx.doi.org/10.1126/science.aar4851.

Song, T., Fei, Z., Yankowitz, M., Lin, Z., Jiang, Q., Hwangbo, K., Zhang, Q., Sun, B., Taniguchi, T., Watanabe, K., et al. (2019). Switching 2D magnetic states via pressure tuning of layer stacking. Nat Mater *18*, 1298-1302. http://dx.doi.org/10.1038/s41563-019-0505-2.

Sun, Z., Yi, Y., Song, T., Clark, G., Huang, B., Shan, Y., Wu, S., Huang, D., Gao, C., Chen, Z., et al. (2019). Giant nonreciprocal second-harmonic generation from antiferromagnetic bilayer $CrI_3$. Nature *572*, 497-501. http://dx.doi.org/10.1038/s41586-019-1445-3.

Verzhbitskiy, I.A., Kurebayashi, H., Cheng, H., Zhou, J., Khan, S., Feng, Y.P., and Eda,

G. (2020). Controlling the magnetic anisotropy in $Cr_2Ge_2Te_6$ by electrostatic gating. Nat Electron *3*, 460-465. http://dx.doi.org/10.1038/s41928-020-0427-7.

Wang, H., Liu, Y., Wu, P., Hou, W., Jiang, Y., Li, X., Pandey, C., Chen, D., Yang, Q., Wang, H., et al. (2020a). Above Room-Temperature Ferromagnetism in Wafer-Scale Two-Dimensional van der Waals $Fe_3GeTe_2$ Tailored by a Topological Insulator. ACS Nano *14*, 10045-10053. http://dx.doi.org/10.1021/acsnano.0c03152.

Wang, X., Cao, J., Lu, Z., Cohen, A., Kitadai, H., Li, T., Tan, Q., Wilson, M., Lui, C.H., Smirnov, D., et al. (2021). Spin-induced linear polarization of photoluminescence in antiferromagnetic van der Waals crystals. Nat Mater http://dx.doi.org/10.1038/s41563-021-00968-7.

Wang, Y., Deng, L., Wei, Q., Wan, Y., Liu, Z., Lu, X., Li, Y., Bi, L., Zhang, L., Lu, H., et al. (2020b). Spin-valley locking effect in defect states of monolayer $MoS_2$. Nano Lett *20*, 2129-2136. http://dx.doi.org/10.1021/acs.nanolett.0c00138.

Webster, L., and Yan, J.-A. (2018). Strain-tunable magnetic anisotropy in monolayer $CrCl_3$, $CrBr_3$, and $CrI_3$. Physical Review B *98* http://dx.doi.org/10.1103/PhysRevB.98.144411.

Wu, S., Buckley, S., Jones, A.M., Ross, J.S., Ghimire, N.J., Yan, J., Mandrus, D.G., Yao, W., Hatami, F., Vučković, J., et al. (2014). Control of two-dimensional excitonic light emission via photonic crystal. 2D Mater *1*, 011001. http://dx.doi.org/10.1088/2053-1583/1/1/011001.

Wu, S., Buckley, S., Schaibley, J.R., Feng, L., Yan, J., Mandrus, D.G., Hatami, F., Yao, W., Vuckovic, J., Majumdar, A., et al. (2015). Monolayer semiconductor nanocavity lasers with ultralow thresholds. Nature *520*, 69-72. http://dx.doi.org/10.1038/nature14290.

Yang, S., Xu, X., Zhu, Y., Niu, R., Xu, C., Peng, Y., Cheng, X., Jia, X., Huang, Y., Xu, X., et al. (2021). Odd-Even Layer-Number Effect and Layer-Dependent Magnetic Phase Diagrams in MnBi$_2$Te$_4$. Phys Rev X *11*, 011003. http://dx.doi.org/10.1103/PhysRevX.11.011003.

Ye, Y., Wong, Z.J., Lu, X., Ni, X., Zhu, H., Chen, X., Wang, Y., and Zhang, X. (2015a). Monolayer excitonic laser. Nat Photonics *9*, 733-737. http://dx.doi.org/10.1038/nphoton.2015.197.

Ye, Y., Wong, Z.J., Lu, X., Ni, X., Zhu, H., Chen, X., Wang, Y., and Zhang, X. (2015b). Monolayer excitonic laser. Nature Photonics *9*, 733-737. http://dx.doi.org/10.1038/nphoton.2015.197.

Zhang, W.-B., Qu, Q., Zhu, P., and Lam, C.-H. (2015). Robust intrinsic ferromagnetism and half semiconductivity in stable two-dimensional single-layer chromium trihalides. Journal of Materials Chemistry C *3*, 12457-12468. http://dx.doi.org/10.1039/c5tc02840j.

Zhang, X.-X., Li, L., Weber, D., Goldberger, J., Mak, K.F., and Shan, J. (2020). Gate-tunable spin waves in antiferromagnetic atomic bilayers. Nat Mater *19*, 838-842. http://dx.doi.org/10.1038/s41563-020-0713-9.

Zhao, L., Shang, Q., Gao, Y., Shi, J., Liu, Z., Chen, J., Mi, Y., Yang, P., Zhang, Z., Du, W., et al. (2018). High-temperature continuous-wave pumped lasing from large-area monolayer semiconductors grown by chemical vapor deposition. ACS Nano *12*, 9390-

9396. http://dx.doi.org/10.1021/acsnano.8b04511.

Zheng, G., Xie, W.-Q., Albarakati, S., Algarni, M., Tan, C., Wang, Y., Peng, J., Partridge, J., Farrar, L., Yi, J., et al. (2020). Gate-Tuned Interlayer Coupling in van der Waals Ferromagnet $Fe_3GeTe_2$ Nanoflakes. Phys Rev Lett *125*, 047202. http://dx.doi.org/10.1103/PhysRevLett.125.047202.

Zhong, D., Seyler, K.L., Linpeng, X., Cheng, R., Sivadas, N., Huang, B., Schmidgall, E., Taniguchi, T., Watanabe, K., McGuire, M.A., et al. (2017). Van der Waals engineering of ferromagnetic semiconductor heterostructures for spin and valleytronics. Sci Adv *3*, e1603113. http://dx.doi.org/10.1126/sciadv.1603113.

# Acknowledgements

We acknowledge financial support from National Natural Science Foundation of China (51872039, 52021001), Science and Technology Program of Sichuan (M112018JY0025).

# Author contributions

B.P. developed the concept, designed the experiment. B.P., Z.Y.C, L.J.D, D.F.L prepared the manuscript and discussed the mechanism of magnetic encoding. Z.Y.C, Z.L. prepared the samples and performed the PL and Raman measurements. Y.L., contributed to fully relativistic calculations.

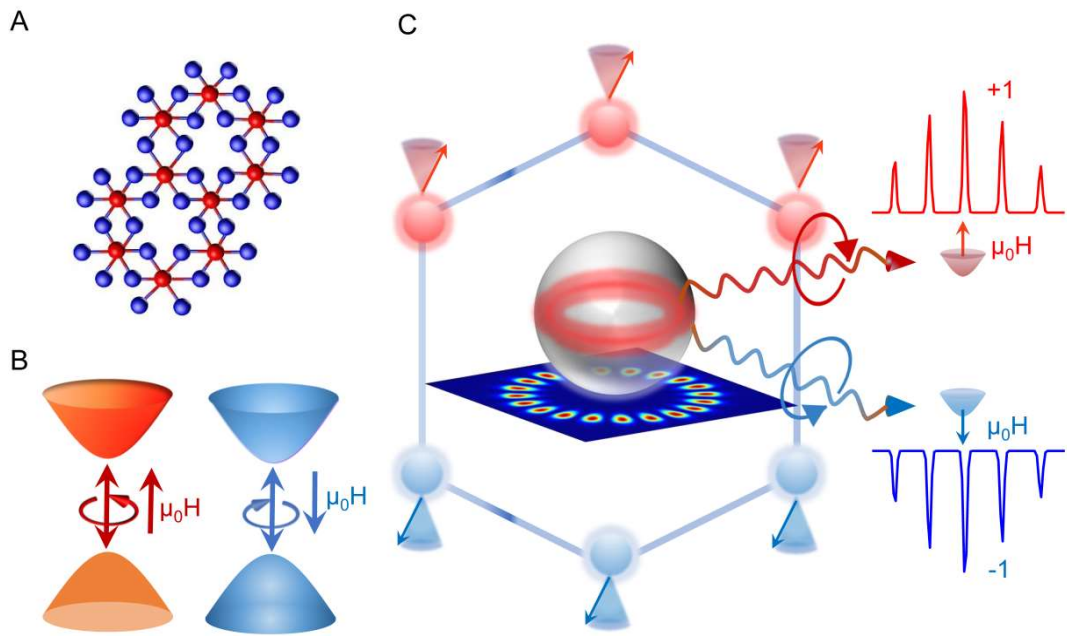

**Fig. 1 CrI$_3$/microsphere WGM microcavities.** (A) Atomic structure of monolayer CrI$_3$. Red and blue balls represent Cr and I atoms, respectively. (B) Schematic of magnetic control of righ-hand (red) and left-hand (blue) circularly polarized PL light. (C) Magnetic encoding of WGM PL in CrI$_3$/microsphere cavities. Strong WGM oscillations lead to narrow circularly polarized WGM PL associated with magnetic orders upon excitation by circular polarization-resolved light.

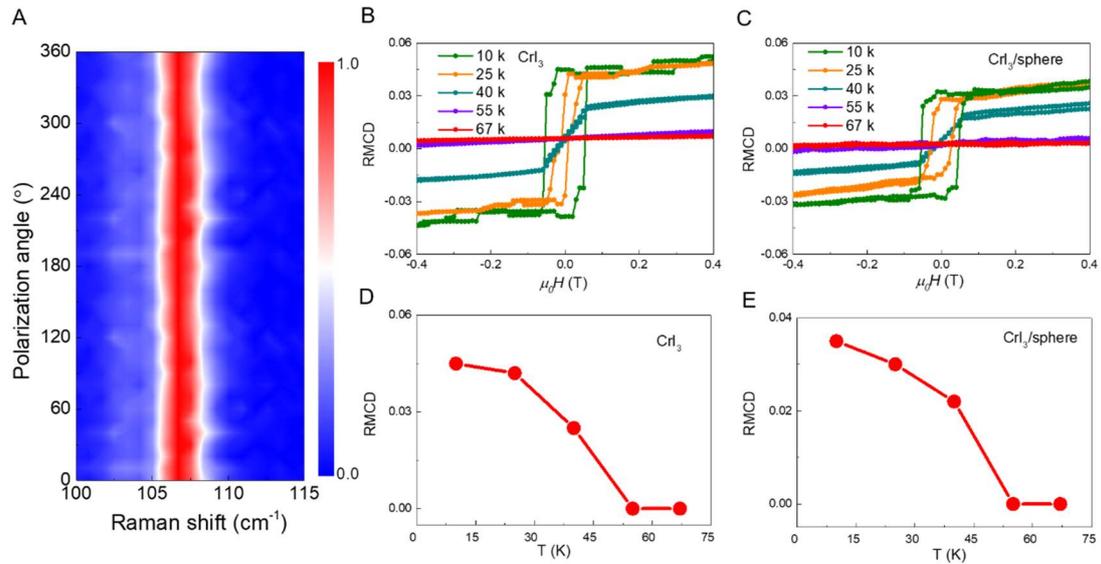

**Fig. 2 Polarization-dependent Raman and RMCD measurements of CrI₃/microsphere and bare CrI₃.** (A) Polarization angle dependent Raman maps of the CrI₃ bulk beneath a 10 μm SiO₂ microsphere. (B, C) Temperature-dependent RMCD measurements of the bare CrI₃ and CrI₃/microsphere (10 μm). (D, E) Extracted RMCD intensities as a function of temperature from (B, C), showing a *Tc* of ~50 K.

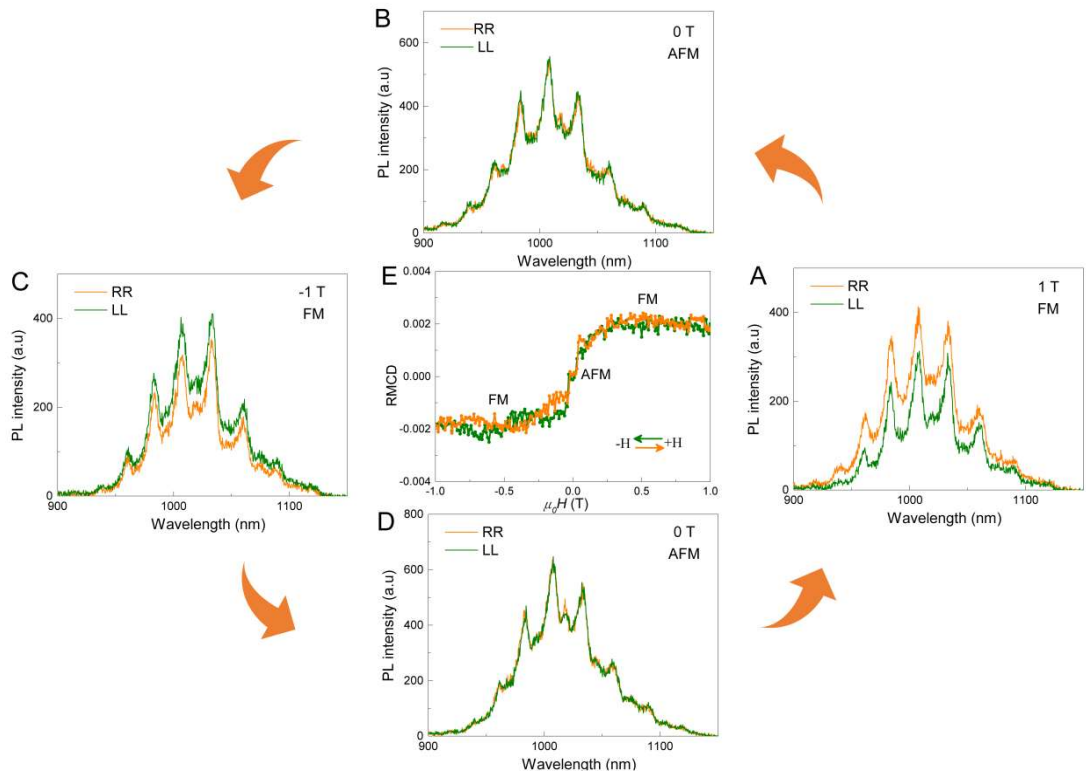

**Fig. 3 Magnetic control of WGM PL of CrI$_3$ beneath a 10 μm SiO$_2$ microsphere.**
(A-D) Circularly polarized PL spectra upon 2 mW 633 nm excitation with magnetic field at +1, 0, -1 and 0 T. (E) Corresponding RMCD curve as a function of magnetic field, which is consistent with the magnetic control of WGM PL.

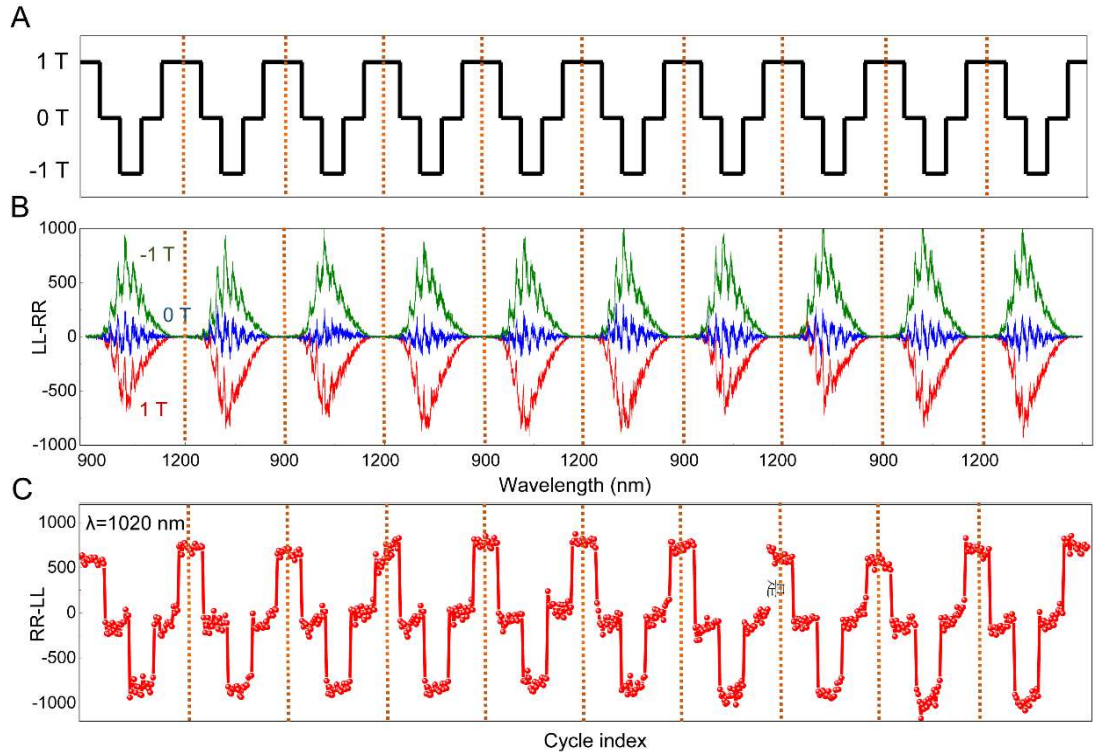

**Fig. 4 Reversible magnetic control and coding of helical luminescence in the CrI$_3$/microsphere.** (A) Schematic of the cycle-dependent magnetic field setup with a sequence of 1 T, 0 T, -1 T, 0 T and 1 T. (B) Recycling control of PL intensity difference between helical WGM PL of CrI$_3$/microsphere upon 2 mW 633 nm excitation by magnetic field. (C) Magnetic control and tri-state coding of WGM PL peaks at 1020 nm extracted from Fig. 4B.

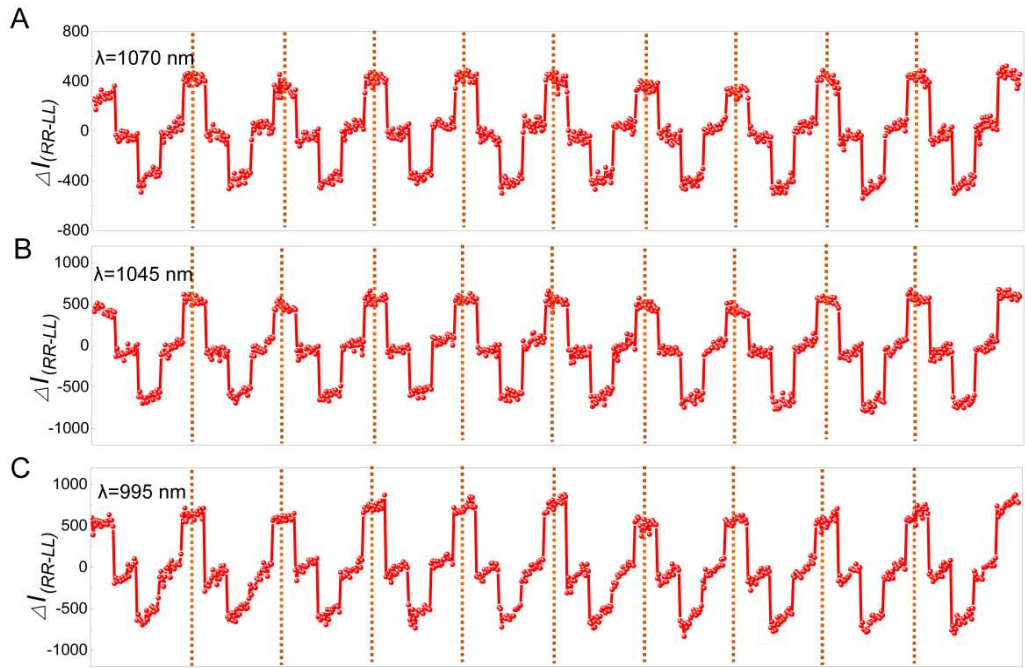

**Fig.5 Multiwavelength magnetic coding through PL intensity difference of helical WGM luminescence.** (A-C) Reversible magnetic control and tri-state coding at 1070, 1045 and 995 nm in selected magnetic fields of +1, 0, -1 T.

.

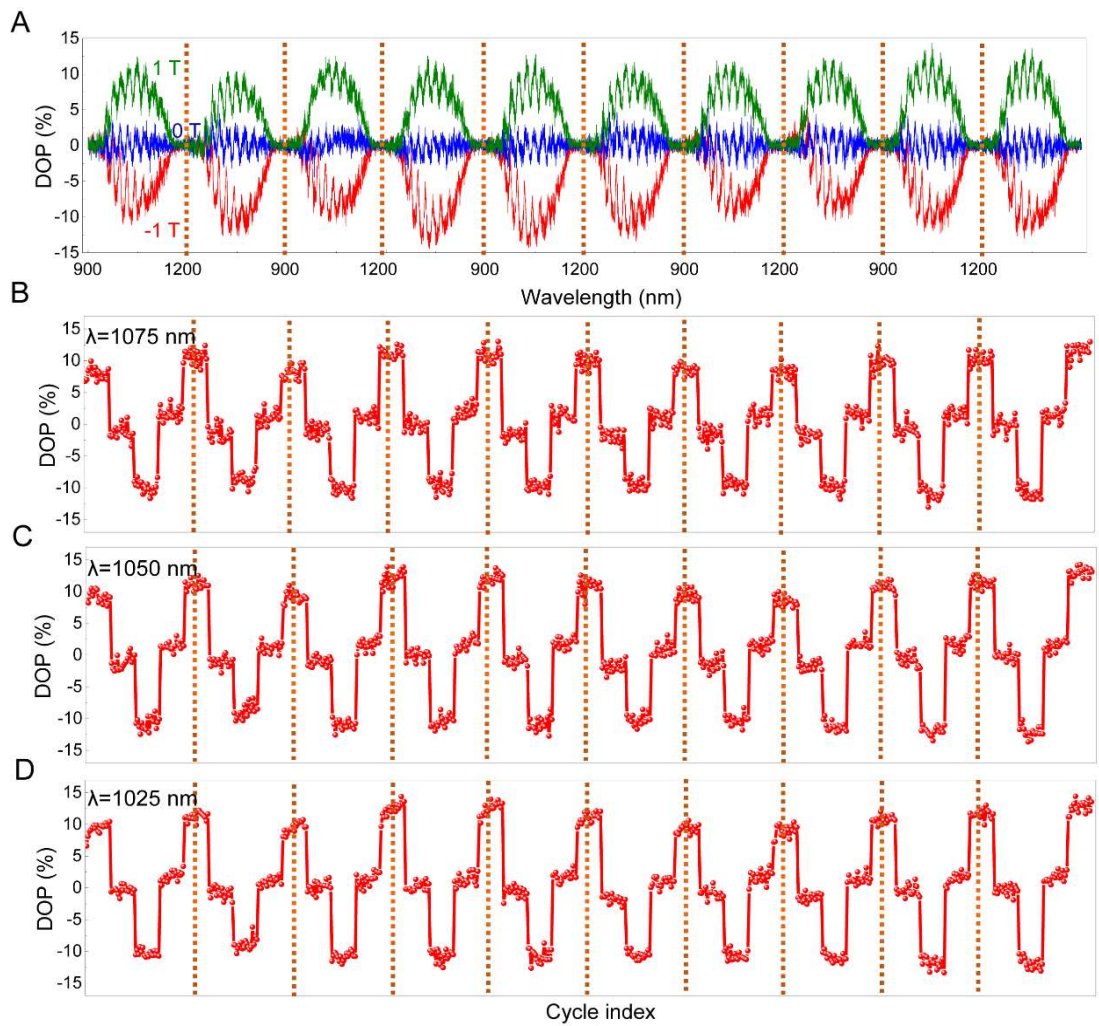

**Fig.6 Multiwavelength magnetic coding through helicity of WGM luminescence. b** (A)Recycling magnetic control of DOP. (B-D) Corresponding tri-state magnetic encoding of DOP at 1075, 1050 and 1025 nm in selected magnetic fields of +1, 0, -1 T.